\documentclass[%
 reprint,
 amsmath,amssymb,
 aps,
prb,
 longbibliography,
floatfix,
 lengthcheck,%
]{revtex4-1}

\usepackage{graphicx}
\usepackage{dcolumn}
\usepackage{bm}
\usepackage{hyperref}

\begin{document}


\title{Observation of isosceles triangular electronic structure around excess iron atoms in Fe$_{1+\delta}$Te}

\author{T. Machida$^{1,2,*}$, K. Kogure$^{1}$, T. Kato$^1$, H. Nakamura$^1$, H. Takeya$^3$, T. Mochiku$^{2}$, S. Ooi$^{2}$, Y. Mizuguchi$^4$, Y. Takano$^{3}$, K. Hirata$^{2}$ and H. Sakata$^{1}$}

\affiliation{$^{1}$Department of Physics, Tokyo University of Science, 1-3 Kagurazaka, Shinjuku-ku, Tokyo 162-8601, Japan\\
$^{2}$Superconducting Properties Unit, National Institute for Materials Science, 1-2-1 Sengen, Tsukuba, Ibaraki 305-0047, Japan \\
$^{3}$Superconducting Wires Unit, National Institute for Materials Science, 1-2-1 Sengen, Tsukuba, Ibaraki 305-0047, Japan \\
$^{4}$Electrical and Electronic Engineering, Tokyo Metropolitan University, 1-1 Minami-osawa, Hachioji, Tokyo 192-0397, Japan}
\date{\today}

\begin{abstract}
We present scanning tunneling microscopy and spectroscopy studies around an individual excess Fe atom, working as a local perturbation, in the parent material of the iron-chalcogenide superconductor Fe$_{1+\delta}$Te.
Spectroscopic imaging reveals a novel isosceles triangular electronic structure around the excess Fe atoms.
Its spatial symmetry reflects the underlying bicollinear antiferromagnetic spin state and the structural monoclinic symmetry.
These findings provide important clues to understand the role of the excess Fe atoms, which complicate the understanding of the phenomena occurring in iron-chalcogenide materials.
\end{abstract}

\pacs{Valid PACS appear here}
\maketitle
\section{Introduction}
A symmetry-breaking ground state is often a crucial key for high-temperature superconductivity.
Such symmetry-breaking states emerge in the parent phase of the iron-based superconductors: electronic\cite{Kasahara,TMChuang,Allan,TMachida_1,XZhou,JHChu}, spin\cite{Cruz,QHuang,SLi,WBao} and orbital\cite{MYi} states with $C_{2}$ symmetry, which are associated with the breaking of structural $C_{4}$ rotational symmetry.
Establishment of the physical picture that lies behind these states is paramount in order to unveil the mechanism of the superconductivity.
However, the complex entanglement of the electronic, spin, orbital, and lattice degrees of freedom obscures the true nature of this exotic parent state.

Useful insight into this problem can be gained by visualizing the electronic structure around a local perturbation, which can be achieved using scanning tunneling microscopy (STM).
Notably, in strongly correlated systems where several kinds of order emerge, the electronic structure around local perturbations reflects the spatial symmetry of the underlying order. For example, local perturbations in the cuprate superconductors, which include impurities\cite{SHPan,TMachida_2,Hudson} and vortices\cite{Matsuba,Levy}, create local density-of-states patterns with four-fold symmetry that reflect the symmetry of the superconducting (SC) order parameter around the local perturbations.

The use of local perturbations in the iron-based superconductors is also an intriguing approach to determine their pairing symmetries in the SC state\cite{Hanaguri_1,CLSong,Hanaguri_2} and to capture the spatial symmetry of the electronic structure in the parent state\cite{TMChuang,Allan,TMachida_1,XZhou}.
Particularly, in the parent state of Ca(Fe$_{1-x}$Co$_{x}$)$_{2}$As$_{2}$ the electronic structure around a Co impurity exhibits a pattern with $C_{2}$ symmetry that is elongated along the spin antiparallel direction of the underlying collinear antiferromagnetic (AFM) order (i.e., the crystal $a$-axis)\cite{TMChuang,Allan}.
Theories have predicted that this $C_{2}$ symmetric structure is intimately related to the underlying AFM\cite{HHuang,TZhou} and orbital\cite{Inoue} ordering.
This implies that the observation of $C_{2}$-symmetric electronic structure around an impurity might provide important clues to enable understanding of the $C_{2}$-symmetric phenomena that occur in the parent state.
Thus, the application of local perturbations to other iron-based superconductors with different spin, orbital, and crystal lattice symmetries promises to open up a new avenue to resolving the highly controversial issues associated with the physical picture lying behind the $C_{2}$ symmetric phenomena in the parent materials.
\begin{figure}[htb]
\begin{center}
\includegraphics[width=7cm]{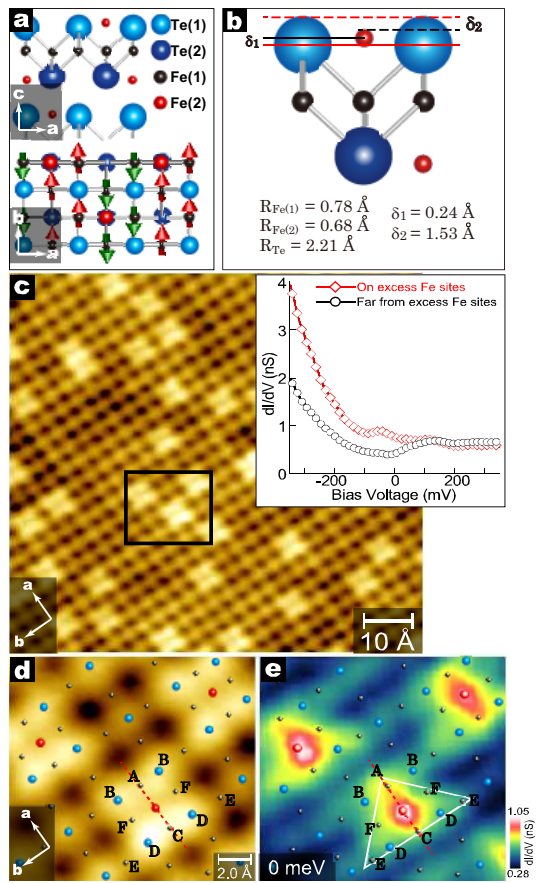}
\end{center}
\caption{
(Color online) \textbf{a} and \textbf{b}, Schematic illustrations of the crystal structure of Fe$_{1+\delta}$Te.
In these figures, light blue, dark blue, black, and red spheres represent Te atoms above the Fe layer [Te(1)], Te atoms below the Fe layer [Te(2)], Fe atoms in the Fe layer [Fe(1)], and excess interstitial Fe atoms [Fe(2)], respectively.
In \textbf{b}, the difference between the atomic heights of Te(1) and Fe(2) from Fe(1) layer ($\delta_{1}$) is determined by using the previous neutron diffraction measurement on Fe$_{1.05}$Te at 5 K in Ref. 33. The difference between the ionic radii of Te(1) and Fe(2) is shown as $\delta_{2}$ (see text). $R_{\mathrm{Fe(1)}}$, $R_{\mathrm{Fe(2)}}$, and $R_{\mathrm{Te}}$ are the ionic radii of Fe(1), Fe(2), and Te atoms, respectively.
\textbf{c}, STM image taken at $V_{\mathrm{B}}$ = +500 mV and $I_{\mathrm{set}}$ = 500 pA on an 86 \AA\ $\mathrm{\times}$ 86 \AA\ field of view (FOV).
The white arrows indicate the crystal $a$- and $b$-axes.
The inset shows tunneling spectra averaged over an area far from the excess Fe atoms (black circles) and over ten Fe(2) sites (red diamonds).
\textbf{d}, Magnified STM image for the 15.8 \AA\ $\mathrm{\times}$ 14.5 \AA\ FOV indicated by the black box in \textbf{c}.
\textbf{e}, Zero-bias conductance map [G(r, 0 meV)] for the same FOV as in \textbf{d}.
In \textbf{d} and \textbf{e}, the light blue, black, and red spheres correspond to Te(1), Fe(1), and Fe(2) atoms, respectively.
Atomic sites near excess Fe atoms are labelled "A" to "F".
}
\label{fig1}
\end{figure}
Here we focus on the parent material of the iron-chalcogenide superconductors, Fe$_{1+\delta}$Te.
In this material, a structural transition from tetragonal to monoclinic symmetry occurs at a temperature $T_{\mathrm{s}}$.
This crystal symmetry breaking coincides with the onset of bicollinear-type AFM order (Fig. \ref{fig1}\textbf{a})\cite{SLi,WBao}.
In addition, Fe$_{1+\delta}$Te has the simplest crystal structure consisting of only an Fe-Te tetrahedral network in which a small number of excess interstitial Fe atoms act as a local perturbation, as shown in Fig. \ref{fig1}\textbf{a} and \textbf{b}.
This structural simplicity removes the necessity of taking into account the effects of other layers on the electronic structure, as is required for other iron-based superconducting materials\cite{XZhou,Nishizaki}.
In this paper, we present spectroscopic-imaging studies around an individual excess Fe atom, working as a local perturbation, in Fe$_{1+\delta}$Te.
The results indicate that a novel isosceles triangular electronic structure emerges around the excess Fe atom near Fermi energy.
The spatial symmetry of the observed isosceles triangular structure reflects the spin configuration of the bicollinear AFM order and the structural monoclinic symmetry.
These findings will be a crucial key to understand the role of the excess Fe atom, which obscures the physical picture of the phenomena occurring in the iron-chalcogenide SCs.

\section{Experiments}
Single crystals of Fe$_{1.07}$Te were grown by simple melting growth technique \cite{TMachida_1}.
The structural transition and AFM transition temperatures were determined by the magnetization and resistivity measurements to be 64 K.
We use a laboratory-built cryogenic STM that operates at 4.2 K in a pure helium gas environment.
Samples were cleaved in situ at 4.2 K.
An electrochemically polished Au wire was used as an STM tip.
Before proceeding to measurements on a sample, we scanned on
an Au thin film (200 nm thickness) deposited on a cleaved surface of a mica to verify the quality of the STM tip.
The STM topographic images were obtained in constant-current mode.
The $dI/dV$ conductance spectra were obtained by numerical differentiation of the $I$-$V$ characteristics measured at each location.
The orientations of the crystal $a$- and $b$-axes were determined by thorough Fourier transform analyses of the topographic images on fields of view (FOV) larger than 400 $\mathrm{\times}$ 400 \AA\ (details are given elsewhere\cite{TMachida_1}).
\begin{figure}[htb]
\begin{center}
\includegraphics[width=6cm]{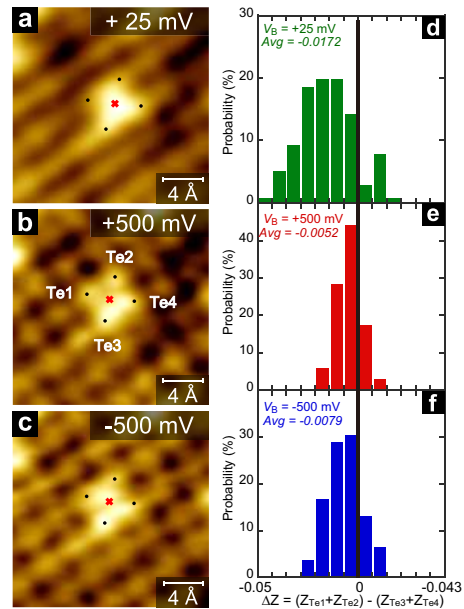}
\end{center}
\caption{(Color online)
\textbf{a} to \textbf{c} STM images taken at $V_{\mathrm{B}}$ = +25, +500, and -500 mV, respectively. \textbf{d} to \textbf{f} Histograms of the values of $\Delta Z$ = ($Z_{1}$ + $Z_{2}$) - ($Z_{3}$ + $Z_{4}$) at $V_{\mathrm{B}}$ = +25, +500, and -500 mV, respectively, where $Z_{i}$ ($i$ = 1,2,3,4) is the STM signal at the Te(1)$_{i}$ site designated in \textbf{b}.
}
\label{fig2}
\end{figure}
\begin{figure}[htb]
\begin{center}
\includegraphics[width=7cm]{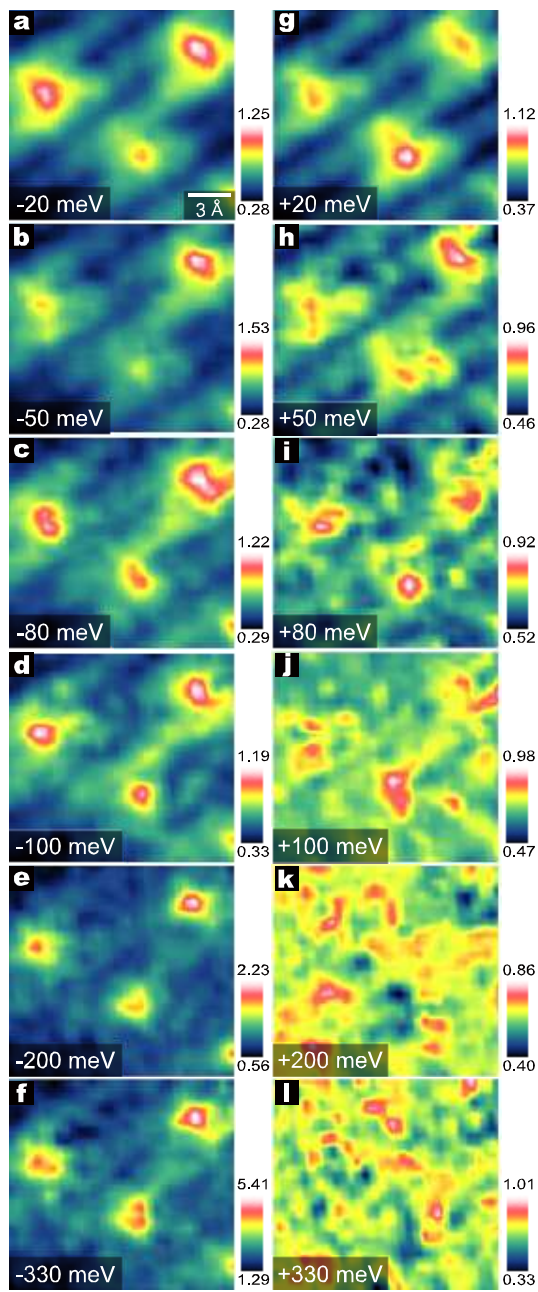}
\end{center}
\caption{
(Color online)
\textbf{a}-\textbf{f}, Conductance maps taken at energies from -20 to -330 meV.
\textbf{g}-\textbf{h}, Conductance maps taken at energies from +20 to +330 meV.
All maps were taken under the same conditions ($V_{\mathrm{B}}$ = +500 mV and $I_{\mathrm{set}}$ = 500 pA) and on the same FOV as in Figs. 1\textbf{d} and \textbf{e}.
}
\label{fig3}
\end{figure}

\section{Results}
\subsection{A typical STM image around an excess Fe atom: A four-leaf clover shape}
A typical topographic image contains randomly distributed bright spots corresponding to the excess Fe atoms [Fe(2)] and an approximately square lattice structure with a period of ~3.8 \AA\ corresponding to the topmost Te atoms [Te(1)], as shown in Fig. \ref{fig1}\textbf{c}.
In the magnified topographic image in Fig. \ref{fig1}\textbf{d}, the patterns around the Fe(2) atoms have "four-leaf clover" shapes characterized by a local minimum at the Fe(2) site and brighter contrast at the four nearest-neighbor Te(1) atoms compared to other Te(1) atoms that do not surround an Fe(2) atom.

These features of the observed "four-leaf clover" shape can be well-explained by considering the relation among the ionic radii and the positions of the Fe(2) and Te atoms\cite{caution}. The local minimum at the excess Fe site appears to be mainly responsible for a difference between the ionic radii of the excess Fe and Te atoms. Even though the Fe(2) height from the Fe(1) atom ($\sim$ 1.99 \AA) is slightly higher than the Te height ($\sim$ 1.75 \AA), the ionic radius of Fe(2) (0.68 \AA\ for divalent and for square planer coordination) is significantly smaller than that of Te (2.21 \AA)\cite{Shannon}.
Thus, the difference of ionic radii $\delta_{2}$ =1.53 \AA\ is grater than that of the atomic heights from Fe(1) atom $\delta_{1}$ = 0.24 \AA, as shown in Fig. \ref{fig1}\textbf{b}.
Consequently, the STM topographic profile $Z(\mbox{\boldmath$r$})$ at the excess Fe atom shows local minimum, since an STM tip follows extent of the electron clouds which is intimately tied to the ionic radius.

We next discuss the brighter contrast at the four Te(1) sites adjacent to the excess Fe atom than other Te(1) sites.
According to the recent X-Ray and neutron diffraction studies \cite{TMachida_A1,EERodriguez}, the Te height from the Fe(1) atom slightly increases with increasing the amount of the excess Fe atoms.
From these diffraction results, it is plausible that the excess Fe atoms make the height of the Te atoms just around the excess Fe atoms higher than those of other Te atoms.

\subsection{Electronic structure around an excess Fe atom: An isosceles triangular structure near $E_{\mathrm{F}}$ }
To explore the effect of the excess Fe on the energy-resolved electronic structure, we compared the tunneling spectra taken at the excess Fe sites and at positions far away from them, as shown in the inset of Fig. 1\textbf{c}.
As previously reported\cite{TMachida_1}, both spectra exhibit large particle-hole asymmetry, which is signified by greater conductance for negative energies than for positive. At the excess Fe sites, the particle-hole asymmetry becomes more pronounced and the conductance near $E_{\mathrm{F}}$ is enhanced.

We first focus on the spatial evolution of the electronic structure around an individual excess Fe atom.
Figure 1\textbf{e} shows a zero-bias conductance map with a spatial resolution of 0.55 \AA\ and the same FOV as in Fig. \ref{fig1}\textbf{d}.
Two novel features are observed.
First, the spatial pattern of the zero-bias conductance has an isosceles triangular shape with the vertex at the "A" site and the corners of the base at the "E" sites.
From the perspective of the spatial symmetry, this pattern has only one mirror plane connecting "A" with "C" sites as shown by dashed lines in Fig. \ref{fig1}\textbf{d} and \textbf{e}.
A comparison between STM images taken at high ($\pm$ 500 mV) and low (+25 mV) bias voltages also supports the existence of this isosceles triangular electronic structure near $E_{\mathrm{F}}$, as shown in Figs. \ref{fig2}\textbf{a}-\textbf{c}.
At high bias voltages ($V_{\mathrm{B}}$ = $\pm$ 500 mV), a "four-leaf clover" shaped pattern is visible.
On the other hand, at a low bias voltage ($V_{B}$ = +25 mV), a triangular pattern appears around the excess Fe atom.
This comparison clearly indicates that the spatial pattern of the low energy LDOS is the triangular shape.

The second novel feature is that the observed isosceles triangular structures direct the preferred orientation.
To analyze statistically the direction of the isosceles triangular structure, we introduced a value of $\Delta Z = (Z_{\mathrm{Te1}}+Z_{\mathrm{Te2}})-(Z_{\mathrm{Te3}}+Z_{\mathrm{Te4}})$, where the $Z_{\mathrm{Te}i}$ ($i = 1, 2, 3, 4$)
is the intensity of the STM image at one of four Te sites adjacent to an excess Fe atom as shown in Fig. \ref{fig2}.
For example, if the energy-integrated LDOS pattern is spatially isotropic around the  excess Fe atom, the value of $\Delta Z$ should be approximately zero, since the topographical signals at the four Te sites are same.
If the energy-integrated LDOS pattern shows an isosceles triangular shape as in Fig. \ref{fig1}\textbf{e}, $\Delta Z$ should be negative, since the energy-integrated LDOS at Te1 or Te2 site is lower than that at Te3 or Te4 site.
If the isosceles triangular LDOS pattern trends to the opposite direction, $\Delta Z$ should be positive.
Figures \ref{fig2}\textbf{d}-\textbf{f} show histograms of $\Delta Z$ with respect to 140 excess Fe atoms.
For $V_{\mathrm{B}}$ = +25 mV at which the spatial pattern is the triangular shape rather than the "four-leaf clover" shape as in Fig. \ref{fig2}\textbf{a}, the $\Delta Z$ values are weighted toward the negative side and the number of the negative $\Delta Z$ reach approximately 90 percent.

\begin{figure}[tb]
\begin{center}
\includegraphics[width=6cm]{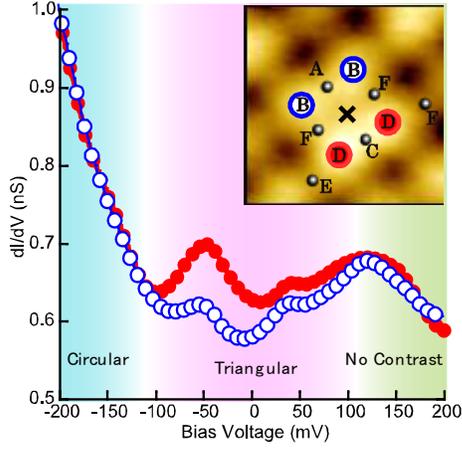}
\end{center}
\caption{(Color online)
In the main panel, the tunnelling spectra averaged over ten "B" (out of the triangular) and "D" (inside) sites as shown in the inset. Inset indicates a typical STM image including a single excess Fe atom.
}
\label{fig4}
\end{figure}
\begin{figure}[tb]
\begin{center}
\includegraphics[width=7cm]{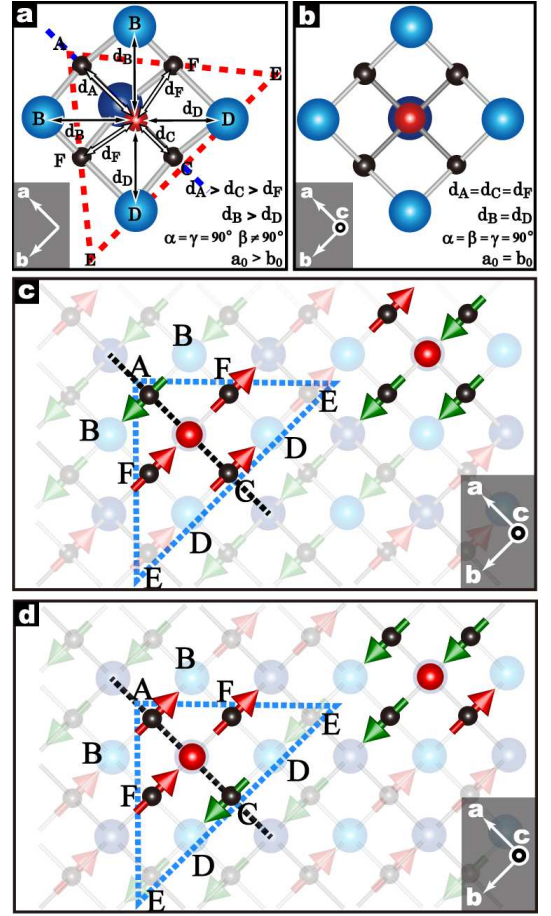}
\end{center}
\caption{
(Color online)
\textbf{a} and \textbf{b}, Schematic pictures of the monoclinic and tetragonal structures viewed along the crystal $c$-axis.
In \textbf{a}, for reasons of clarity the angle between the $a$- and $c$-axes ($\beta$) is exaggerated to 80$^{\circ}$, which is smaller than the actual angle (89.2$^{\circ}$).
The distances between the excess Fe site and the neighboring atomic sites ("A" to "F") are labelled as "$d_{\mathrm{A}}$" to " $d_{\mathrm{F}}$ ".
In the actual crystal structure, $d_{\mathrm{A}}$ = 2.801 \AA\ $>$ $d_{\mathrm{c}}$ = 2.763 \AA\ $>$ $d_{\mathrm{F}}$ = 2.733 \AA, and $d_{\mathrm{B}}$ = 2.707 \AA\ $>$ $d_{\mathrm{D}}$ = 2.702 \AA.
\textbf{c} and \textbf{d}, Spin configurations of the bicollinear AFM order in the Fe(1)-layer.
For clarity, the spins adjacent to the excess Fe(2) atoms are emphasized.
}
\label{fig5}
\end{figure}

\subsection{Energy dependence of the isosceles triangularity}
We mapped the conductance at several energies and investigated the evolution of the observed triangular structure, as shown in Fig. \ref{fig3}.
At negative energies, the electronic structure is clearly triangular from 0 to -50 meV and becomes approximately isotropic below -100 meV.
At positive energies, the triangular pattern is clearly visible from 0 to +50 meV but there is no physically meaningful contrast except for noise above +100 meV.

For more quantitative indication of the energy dependence of the spatial $dI/dV$ pattern around the excess Fe site, we compare the $dI/dV$ spectra at "B" sites and "D" sites, as shown in Fig. \ref{fig4}.
This comparison provides an energy dependence of the isosceles triangularity, since the "B" and "D" sites are located at outside and inside of the isosceles triangular, respectively.
For instance, if the isosceles triangular pattern appears in a $dI/dV$ map, the $dI/dV$ value at "D" sites should be greater than that at "B" sites.
On the other hand, if there is a isotropic pattern or no contrast in a $dI/dV$ map, the difference between the $dI/dV$ at both sites should be zero.
Figure \ref{fig4} shows this comparison, where the spectra are obtained by averaging over ten "B" and "D" sites.
A difference between the spectra is observed in the energy range of $|E| <$ 100 meV and invisible out side of the energy range.
This analysis indicates that the triangular electronic structure resides within the energy range of $|E| <$ 100 meV.

\section{Discussion}
We now discuss the origin of the observed isosceles triangular electronic structure around the excess Fe atoms.
Here, it is important to consider what other features of the material have the same spatial symmetry.
First is the symmetry of the monoclinic crystal structure; the distances from the Fe(2) site to the neighboring atomic sites, as illustrated in Figs. \ref{fig5}\textbf{a} and \textbf{b}.
This crystal monoclinicity makes the hopping integral anisotropic.
The hopping probabilities along the shorter bond directions (from Fe(2) to C, D, or E sites) should be higher than those along the longer bond directions (to A or B sites).
This is consistent with the spatial pattern of the observed triangular electronic structure.

However, the isosceles triangular electronic structure changes into the isotropic below -100 meV as shown in Figs. \ref{fig3} and \ref{fig4}.
This energy evolution of the shape of the LDOS pattern around the excess Fe atom seems to be unexpected from only the monoclinic nature of the crystal.
Therefore, in addition to the monoclinic nature, another factor for the isosceles triangular electronic structure is needed.
It is the spin configuration of the underlying bicollinear AFM order in the vicinity of the excess Fe atoms, as depicted in Fig. \ref{fig5}\textbf{c}.
One of the four spins at the Fe(1) sites surrounding an Fe(2) site is aligned antiparallel to the other three: the spin direction at site "C" is parallel to that at the two "F" sites, whereas the spin direction at site "A" is opposite.
Such spin configuration realizes around all the Fe(2) sites.
This is consistent with the experimental observation that most of the triangular structures point in the same direction.
Although a second spin configuration is possible, as illustrated in Fig. \ref{fig5}\textbf{d}, both configurations are identical with respect to spatial symmetry and we focus here on the configuration shown in Fig. \ref{fig5}\textbf{c}.

If the symmetry of the spin configuration affects predominantly the formation of the triangular electronic structure, the 3$d$ orbital physics of Fe, which is characterized by large Hund's rule coupling ($J_{\mathrm{H}}$), is quite important to understand the origin of the triangular electronic structure.
To explain the metallic parent state with bicollinear AFM order in Fe$_{1+\delta}$Te, several microscopic models have been proposed\cite{Turner,WGYin,JHu,Ducatman}.
In some of these models\cite{Turner,WGYin} it is claimed that Hund's rule coupling between the itinerant electrons and localized spins at each Fe(1) site is imperative for the electron hopping process to occur.
When electrons on Fe(2) atom with large local moments\cite{Blachowski} move to Fe(1) sites on the magnetic background comprised of bicollinear AFM order with double-exchange interactions, the energy gain during this hopping process depends on the angle between the spins at the Fe(2) and Fe(1) sites.
This is because the Hund's rule coupling renders the spin of the hopping electron parallel to that at the Fe(1) site\cite{Anderson}.
Although the spin direction at the Fe(2) site is unknown, the set of hopping probabilities between an Fe(2) site and its neighboring Fe(1) sites has the same spatial symmetry due to the difference in energy gain induced by the spin configuration around the excess Fe atom.

Our results show that the triangular structure is clearly observed within the energy range $|E_{\mathrm{t}}|$ $<$ 50 meV.
If our proposed scenario is valid, $|E_{\mathrm{t}}|$ should be comparable to the energies of the superexchange interactions between neighboring Fe(1) atoms or to $J_{\mathrm{H}}$.
The value of $J_{\mathrm{H}}$ is known to be 0.4 - 0.8 eV\cite{WLYang}, which is considerably higher than $|E_{\mathrm{t}}|$, whereas the superexchange interaction energies determined by inelastic neutron scattering\cite{Lipscombe} are several tens of meV, in agreement with $|E_{\mathrm{t}}|$.
It is noted that these are bulk values and would be modified by around the excess Fe atom due to the strain field induced by the excess Fe itself \cite{XLiu}.

\section{Summary}
In summary, we have investigated the local effect of the excess iron atom on the LDOS in Fe$_{1+\delta}$Te by STM/STS measurements.
The novel isosceles triangular structure in the LDOS lies around the excess Fe atoms from -50 meV to +50 meV.
The spatial symmetry of the observed triangular electronic structure corresponds to those of the monoclinic crystal structure and of the spin structure of the underlying AFM order.
Therefore, both the crystal monoclinicity and the spin configuration around the excess Fe atom seems to be required for the formation of the triangular electronic structure.
Since the shape of the low-energy LDOS breaks $C_{4}$-symmetry around the excess Fe atom, the excess Fe might create the effective scattering potential which also breaks $C_{4}$-symmetry.
This appears to be one of possible origin of the in-plane resistivity anisotropy in Fe$_{1+\delta}$Te \cite{JJiang}.
Thus, our results offer the opportunity (i) to understand the $C_{2}$-symmetric electronic transport, and (ii) to establish a theoretical description of the role of the excess Fe atoms, which complicate understanding of the phenomena occurring in iron-chalcogenide materials.

\section{Acknowledgements}
We gratefully acknowledge studies at Tokyo University of Science by JSPS KAKENHI (12018262).


\begin{thebibliography}{36}
\bibitem{caution} Since the tunnelling current (or STM image) contains both a topographic signal and an energy-integrated LDOS signal, there is a finite contribution of the spatial evolution of the energy-integrated LDOS in the tunnelling current. As seen in the energy dependence of the conductance map, the conductance becomes local maximum at the excess Fe site at low energy. If the energy-integrated LDOS signal affects predominantly the STM image at $V_{\mathrm{B}}$ = +500 mV, the STM image should depict a local maximum at the excess Fe site. However, the STM signal at the excess Fe site shows a local minimum in the actual STM image. Therefore, it is conceivable that the energy-integrated LDOS signal does not affect predominantly the STM signals at the high bias voltage more than $|V_{\mathrm{B}}| = 500$ mV.
\end{thebibliography}
\end{document}